\documentclass[12pt,showkeys,superscriptaddress,nofootinbib]{revtex4}
\usepackage{amsmath,amssymb,euscript,yfonts,psfrag,latexsym,dsfont,graphicx,bbm,color,amstext,wasysym,pdfsync,soul,framed}

\graphicspath{{/},{figures/}}

\newcommand{\cE}{{\mathcal E}}

\newcommand{\bI}{{\mathbb I}}
\newcommand{\cF}{{\mathcal F}}

\newcommand{\mC}{{\mathbb C}}
\newcommand{\mE}{{\mathbb E}}
\newcommand{\cD}{{\mathcal D}}

\newcommand{\cS}{{\mathcal S}}
\newcommand{\cK}{{\mathcal K}}
\newcommand{\fD}{{\mathfrak D}}
\newcommand{\fH}{{\mathfrak H}}
\newcommand{\fM}{{\mathfrak M}}

\newcommand{\cC}{{\mathcal C}}

\newcommand{\bbR}{\mathbb{R}}
\newcommand{\bbC}{\mathbb{C}}
\newcommand{\mD}{{\mathbb D}}
\newcommand{\bbD}{{\mD}}
\newcommand{\pio}{{\pi^o\hspace*{-4.5pt}}}

\newcommand{\trace}{\operatorname{\rm trace}}
\newcommand{\diag}{\mathop{\mathrm{diag}}}

\newcommand{\x}{{\chi}}

\newcommand{\rhoh}{{\hat\phi_0}}

\newcommand{\support}{{\rm Supp}}

\definecolor{llgrey}{rgb}{0.9,0.9,0.9}
\definecolor{lgrey}{rgb}{0.6,0.6,0.6}
\definecolor{lred}{rgb}{0.9,0.7,0.7}

\newtheorem{thm}{Theorem}
\newtheorem{conj}{Conjecture}
\newtheorem{cor}{Corollary}
\newtheorem{lemma}{Lemma}

\newtheorem{problem}{Problem}
\newtheorem{remark}{Remark}
\newtheorem{defn}{Definition}

\begin{document}
\title{Positive contraction mappings for classical and quantum Schr\"{o}dinger systems}
\author{Tryphon T. Georgiou}\email{tryphon@umn.edu}\affiliation{University of Minnesota}
\author{Michele Pavon}\email{pavon@math.unipd.it}\affiliation{University of Padova}

\begin{abstract}
{\bf Abstract.} The classical Schr\"{o}dinger bridge seeks the most likely probability law for a diffusion process, in path space, that matches marginals at two end points in time; the likelihood is quantified by the relative entropy between the sought law and a prior, and the law dictates a controlled path that abides by the specified marginals.
Schr\"odinger proved that the optimal steering of the density between the two end points is effected by a multiplicative  functional transformation of the prior; this transformation represents an automorphism on the space of probability measures and has since been studied by Fortet, Beurling and others.
A similar question can be raised for processes evolving in a discrete time and space as well as for processes defined over non-commutative probability spaces.
The present paper builds on earlier work by Pavon and Ticozzi and begins with the problem of steering a Markov chain between given marginals. Our approach is based on the Hilbert metric and leads to an alternative proof which, however, is constructive. More specifically, we show that the solution to the Schr\"{o}dinger bridge is provided by the fixed point of a contractive map. 
We approach in a similar manner the steering of a quantum system across a quantum channel. We are able to establish existence of quantum transitions that are multiplicative {functional} transformations of a given Kraus map, but only for the case of uniform marginals.
As in the Markov chain case, and for uniform density matrices, the solution of the quantum bridge can be constructed from the fixed point of a certain contractive map.
For arbitrary marginal densities, extensive numerical simulations indicate that iteration of a similar map leads to fixed points from which we can construct a quantum bridge. For this general case, however, a  proof of convergence remains elusive.
\end{abstract}

\maketitle
\section{Introduction}

In 1931 Erwin Schr\"odinger published a manuscript on ``the reversal of the laws of nature'' (``\"Uber die Umkehrung der Naturgesetze'').  In it, he raised the following ``new and unorthodox'' \cite{beurling1960automorphism} question regarding Brownian motion. Suppose that the density of Brownian particles is observed at two points in time, $t_0$ and $t_1$, and that the two end-point densities differ from the initial and final marginals of the prior path-space distribution. Schr\"odinger then asked for the most likely random evolution that the particles have taken so as to reconcile the observed  ``improbable but still possible outcome.'' In modern probabilistic language, as observed by F\"{o}llmer some fifty years later \cite{foellmer1988randomfields}, Schr\"{o}dinger was posing (and, to some extent, solving) a problem of large deviations of the empirical distribution. He was working in an abstract setting, although the very foundations of probability theory were still missing! The solution of the large deviations problem requires, in view of Sanov's theorem \cite{sanov1957largedeviations}, solving a maximum entropy problem. 
Schr\"{o}dinger's 1931/32 papers  were followed soon afterwards by works by N. Kolmogoroff on ``the reversibility of the statistical laws of nature'' (``Zur Umkehrbarkeit der statistischen Naturgesetze'') and by Fortet, Beurling and many others on the mathematical issues that
Schr\"odinger's paper raised.

In the present paper, following Pavon and Ticozzi \cite{pavon2010discrete,ticozzi2010time},
we consider discrete-time and discrete-space {classical evolutions} as well as quantum Markovian evolutions. More precisely, we first consider {discrete random vectors}  with given prior distribution and endpoint marginals. We derive a constructive proof of the existence of multiplicative {functional} transformations of the prior {initial-final time joint distribution} that allows connecting the given end-point marginals. The unique solution is in fact the closest law to the prior in a relative entropy sense amongst all probability laws that are in agreement with the two marginals. A key concept is that of the Hilbert metric --this is a metric which is suitable for quantifying distances in homogeneous positive spaces. A similar approach allows a constructive proof for matching uniform marginal density matrices via a multiplicative {functional} transformation of any given prior Kraus map. In essence, this result extends to the non-commutative case a result of Sinkhorn  \cite{sinkhorn1974diagonal,Sinkhorn1964} that any strictly positive stochastic matrix can be transformed into a doubly stochastic matrix via a multiplicative {functional} transformation. Thus, for the quantum case, we establish that any strictly positive Kraus map can be transformed into a doubly stochastic quantum map via a multiplicative {functional} transformation.
{Further, extensive simulations have convinced the authors that the approach works in complete generality, i.e., when the specified marginal densities are not necessarily uniform. However, a rigorous proof as well as a variational principle, in analogy with the classical case, is not available at present.}

The paper is structured as follows. {Section \ref{sec:hilbert} provides an exposition of the Hilbert metric. The corresponding geometry is key in studying the Schr\"odinger bridge in the classical and in the quantum case, in Sections \ref{sec:classical} and \ref{sec:quantum}, respectively. More specifically,
Sections \ref{discreterandom}-\ref{reduction} explain Schr\"odinger's bridges for Markov chains.  Then, in Section \ref{solonestep} the Hilbert metric is used to provide a constructive solution to Schr\"odinger's bridge problem for Markov chains. Section \ref{sec:quantumchannels} overviews the formalism of quantum mechanics followed by a description of a quantum analog of the Schr\"odinger bridge problem in Section \ref{sec:quantumbridge}. Section \ref{sec:doublystochastic} presents a solution of the quantum Schr\"odinger bridge for the special case of uniform marginals. The mathematical statement for this special case represents a generalization of a result of Sinkhorn on the existence of doubly stochastic maps to a corresponding quantum probabilistic analog. This result is followed by a discussion and a conjecture about the general quantum Schr\"odinger problem, namely, that a fixed point of a certain map which is used to construct doubly stochastic maps, suitably modified, has a fixed point for general marginal density matrices as well.
}

\section{The Hilbert metric}\label{sec:hilbert}

This metric was introduced by David Hilbert in 1895 \cite{hilbert1895gerade} while exploring the foundations of geometry. Earlier special cases, its importance and several subsequent developement are being discussed by Bushell \cite{bushell1973hilbert}. More recently, the underlying geometry has proven timely on a range of problems in the study of communication and computations over networks (see \cite{tsitsiklis1986distributed} and in particular the work of Sepulchre and collaborators \cite{sepulchre2010consensus,sepulchre2011contraction} on consensus in non-commutative spaces, as well as the references therein) {and in quantum information theory \cite{reeb2011hilbert}}. A recent survey on the applications in analysis is \cite{lemmens2013birkhoff}. A key result that enables the metric to be used for establishing existence of solutions to various equations was {proved} by Garrett Birkhoff  in 1957 \cite{birkhoff1957extensions}. Following \cite{sepulchre2010consensus}, we highlight the basic elements of the theory.

Let $\cS$ be a real Banach space and let $\cK$ be a closed solid cone in $\cS$, i.e., $\cK$ is closed with nonempty interior and is such that $\cK+\cK\subseteq \cK$, $\cK\cap -\cK=\{0\}$ as well as $\lambda \cK\subseteq \cK$ for all $\lambda\geq 0$. Define the partial order
\[
x\preceq y \Leftrightarrow y-x\in\cK,
\]
and for $x,y\in\cK\backslash \{0\}$, define
\begin{eqnarray*}
M(x,y)&:=&\inf\, \{\lambda\,\mid x\preceq \lambda y\}\\
m(x,y)&:=&\sup \{\lambda \mid \lambda y\preceq x \}.
\end{eqnarray*}
Then, the Hilbert metric is defined on $\cK\backslash\{0\}$ by
\[
d_H(x,y):=\log\left(\frac{M(x,y)}{m(x,y)}\right).
\]
Strictly speaking, it is a {\em projective} metric since it remains invariant under scaling by positive constants, i.e.,
$d_H(x,y)=d_H(\lambda x,y)=d_H(x,\lambda y)$ for any $\lambda>0$ and, thus, it actually measures distance between rays and not elements.

Birkhoff's theorem, which was originally stated for the linear case and suitably extended by Bushell \cite{bushell1973hilbert}, provides bounds on the induced gain of positive maps. More specifically, a map $\cE$ from $\cS$ to $\cS$ is said to be {\em positive} provided it takes the interior of $\cK$ into itself, i.e.,
\[
\cE\;:\;\cK\backslash\{0\} \to \cK\backslash\{0\}.
\]
For such a map define its {\em projective diameter}
\begin{eqnarray*}
\Delta(\cE):=\sup\{d_H(\cE(x),\cE(y))\mid x,y\in \cK\backslash\{0\}\}
\end{eqnarray*}
and the {\em contraction ratio}
\begin{eqnarray*}
\|\cE\|_H:=\inf\{\lambda \mid d_H(\cE(x),\cE(y))\leq \lambda d_H(x,y),\mbox{ for all }x,y\in\cK\backslash\{0\}\}.
\end{eqnarray*}
The Birkhoff-Bushell theorem states the following.

\begin{thm}[\cite{{birkhoff1957extensions},bushell1973hilbert,bushell1973projective}]\label{BBcontraction}
{Let $\cE$ be a positive map as above. If $\cE$ is monotone and homogeneous of degree $m$,
i.e., if
\[
x\preceq y \Rightarrow \cE(x)\preceq \cE(y),
\]
and
\[
\cE(\lambda x)=\lambda^m \cE(x),
\]
then it holds that}
\[
\|\cE\|_H\leq m.
\]
For the special case where $\cE$ is also linear, the (possibly stronger) bound
\[
\|\cE\|_H=\tanh(\frac{1}{4}\Delta(\cE))
\]
also holds.
\end{thm}
Birkoff's result provides a far-reaching generalization of the celebrated Perron-Frobenius theorem \cite{birkhoff1962Perron-Frobenius}. Various other applications of the Birkhoff-Bushell result have been developed such as to positive integral operators and to positive definite matrices \cite{bushell1973hilbert, lemmens2013birkhoff}. 
We will use Theorem \ref{BBcontraction} to establish existence of solutions to certain equations involving Markovian evolutions on a finite time interval. More specifically, $\cS$ will either be $\bbR^n$ or the space of symmetric/Hermitian $n\times n$ matrices with elements in $\bbR$ or $\bbC$, accordingly. Then $\cK$ will either be the positive orthant of vectors with non-negative entries or the cone of non-negative definite matrices, respectively. The former setting will be brought in to give an independent proof of existence of the solution to the Schr\"{o}dinger bridge problem for Markov chains. The latter will be called in for studying {\em quantum operations} (Kraus maps) in Section \ref{sec:quantum}.

\section{Schr\"{o}dinger's problem for discrete random vectors}\label{sec:classical}

Following quite closely Schr\"{o}dinger's original derivation, we give below a rather self-contained presentation for the purpose of later reference and comparison when we deal with the more complex quantum case. First we discuss general discrete random vectors and then ``time windows" of a Markov chain. A paper dealing with the discrete time, continuous state space setting is \cite{beghi1996relative}. Schr\"{o}dinger bridges for Markov chains have been discussed in \cite{pavon2010discrete}. A nice survey with extensive bibliography for the diffusion case is \cite{wakolbinger1992bridges}.
\subsection{Discrete random vectors}\label{discreterandom}

Given a finite\footnote{{The case of a finite state space $\mathcal X$ is chosen for simplicity of exposition. Results extend in a straightforward way to the case of a countable $\mathcal X$.}} set $\mathcal X=\{1,\ldots,N\}$ , we are concerned with probability distributions $P$ on ``trajectories" $x=(x_0,x_1,\ldots,x_T)$ in $\mathcal X^{T+1}$. We write  $\mathcal P$ for the simplex of all such distributions. Let us introduce the {\em coordinate mapping process} $X=\{X(t), 0\le t\le T\}$ by $X(t)(x)=x_t$. {For brevity, we often write $X(t)=x_t$ instead of $X(t)(x)=x_t$.} For $P\in \mathcal P$, we denote by
\[
p(s,x_s;t,x_t):=P(X(t)=x_t \mid X(s)=x_s), \quad 0\le s < t\le T, \quad x_s, x_t\in{\cal X}.
\]
its transition probabilities. We also use $p$, but indexed, to denote marginals. Thus,
\[
p_{t}(x_t):=P(X(t)=x_t),
\]
and similarly, for two-time marginals,
\[
p_{st}(x_s,x_t):=P(X(s)=x_s, X(t)=x_t).
\]
Schr\"{o}dinger's original formulation was set in continuous time and {with continuous state space}. The formulation herein represents the case where Schr\"{o}dinger's problem has undergone  ``coarse graining" in Boltzmann's style in its phase space and where time has also been discretized. Thus, the {\em a priori} model is now given by a  distribution $P\in\mathcal P$ and suppose that in experiments an initial and a final marginal $\mathbf p_0$ and $\mathbf p_T$, respectively, have been observed that differ from the marginals $ p_0$ and $p_T$ of the prior distribution $P$. We denote by ${\mathcal P}(\mathbf p_0,\mathbf p_T)\subset \mathcal P$ the family of distributions having the observed marginals and seek a distribution in ${\mathcal P}(\mathbf p_0,\mathbf p_T)$ which is close to the given prior $P$. Large deviation reasoning \cite{sanov1957largedeviations} requires that we employ as ``distance'' the {\em relative entropy}:

\begin{defn}{\em Let $P,Q\in \mathcal P$, that is, they belong to the simplex  of probability distributions on ${\cal X}^{T+1}$, and let $x=(x_0,x_1,\ldots,x_T)$. If $P(x)=0\Rightarrow Q(x)=0$, we say that the {\em support} of $Q$ is contained in the support of $P$ and write \[\support (Q)\subseteq \support (P).\]
The {\em Relative Entropy} of $Q$ from $P$ is defined to be
\begin{equation}\label{KLdist}\bbD(Q\|P)=\left\{\begin{array}{ll} \sum_{x\in{\cal X}^{T+1}}Q(x)\log\frac{Q(x)}{P(x)}, & \support (Q)\subseteq \support (P),\\
+\infty , & \support (Q)\not\subseteq \support (P).\end{array}\right.,
\end{equation} 
where, by definition,  $0\cdot\log 0=0$.}
\end{defn}

The relative entropy is also known as the {\em information} or {\em Kullback-Leibler divergence}. {As is well known \cite{cover_thomas2006information}, $\bbD(Q\|P)\ge 0$ and $\bbD(Q\|P)=0$ if and only if $Q=P$. }Given this notion of distance,
we seek
a probability law $Q^\circ\in {\mathcal P}(\mathbf p_0,\mathbf p_T)$
which is closest to the prior distribution $P$ in this sense. That is,
we seek a solution to the following problem.
\begin{problem}\label{prob:optimization}{\em Assume that $p(0,\cdot\,; T,\cdot)$ is everywhere positive on its domain. Determine
 \begin{eqnarray}\label{eq:optimization}
Q^\circ={\rm argmin}\{ \bbD(Q\|P)&\mid& Q\in {\mathcal P}(\mathbf p_0,\mathbf p_T)
\}.\nonumber
\end{eqnarray}}
\end{problem}
It turns out that if there is at least one $Q$ in
$\mathcal P(\mathbf p_0,\mathbf p_T)$ such that
$\bbD(Q\|P)<\infty$,
there exists a unique minimizer $Q^\circ$  called
{\em the Schr\"{o}dinger bridge} from $\mathbf p_0$ to $\mathbf p_T$ over $P$.
Now let 
$$Q_{x_0,x_T}=Q\left[\,\cdot\,|X(0)=x_0,X(T)=x_T\right]$$
be the disintegration of Q with respect to the initial and final positions. Then, we have
$$Q(x_0,x_1,\ldots,x_T)=Q_{x_0,x_T}(x_0,x_1,\ldots,x_T)q_{0T}(x_0,x_T),
$$
where we have assumed that $q_{0T}$ is everywhere positive on $\mathcal X\times \mathcal X$.
We get
\begin{equation}
\bbD(Q\|P)=\sum_{x_0x_T}q_{0T}(x_0,x_T)\log \frac{q_{0T}(x_0,x_T)}{p_{0T}(x_0,x_T)}+\sum_{x\in{\cal X}^{T+1}}Q_{x_0,x_T}(x)\log \frac{Q_{x_0,x_T}(x)}{P_{x_0,x_T}(x)} q_{0T}(x_0,x_T).
\end{equation}
This is the sum of two nonnegative quantities. The second becomes zero if and only if $Q_{x_0,x_T}(x)=P_{x_0,x_T}(x)$ for all $x\in{\cal X}^{T+1}$.  Thus, $Q^\circ_{x_0,x_T}(x)=P_{x_0,x_T}(x)$. As already observed by Schr\"{o}dinger, Problem \ref{prob:optimization} then reduces to minimizing
\begin{equation}\bbD(q_{0T}\|p_{0T})=\sum_{x_0x_T}p_{0T}(x_0,x_T)\log \frac{q_{0T}(x_0,x_T)}{p_{0T}(x_0,x_T)}
\end{equation}
with respect to $q_{0T}$ subject to the (linear) constraints
\begin{eqnarray}\sum_{x_T}q_{0T}(x_0,x_T)&=&\mathbf p_0(x_0),\quad x_0\in{\cal X},
\\
\sum_{x_0}q_{0T}(x_0,x_T)&=&\mathbf p_T(x_T),\quad x_T\in{\cal X}.
\end{eqnarray}
The Lagrangian function has the form
\begin{eqnarray}\nonumber&&{\cal L}(q_{0T})=\sum_{x_0x_T}q_{0T}(x_0,x_T)\log \frac{q_{0T}(x_0,x_T)}{p_{0T}(x_0,x_T)}\\&&+\sum_{x_0}\lambda(x_0)\left[\sum_{x_T}q_{0T}(x_0,x_T)-\mathbf p_0(x_0)\right]+\sum_{x_T}\mu(x_T)\left[\sum_{x_0}q_{0T}(x_0,x_T)-\mathbf p_T(x_T)\right].\nonumber
\end{eqnarray}
Setting the first variation equal to zero, we get the (sufficient) optimality condition
$$1+\log q_{0T}^\circ(x_0,x_T)-\log p(0,x_0; T,x_T)-\log p_0(x_0)+\lambda(x_0)+\mu(x_T)=0,
$$
where we have used the expression $p_{0T}(x_0,x_T)=p_0(x_0)p(0,x_0;T,x_T)$. Hence,  the ratio $q_{0T}^\circ(x_0,x_T)/p(0,x_0; T,x_T)$ factors into a function of $x_0$ times a function of $x_T$; {these are denoted $\hat{\varphi}(x_0)$ and $\varphi(x_T)$, respectively.} We can then write the optimal $q_{0T}^\circ(\cdot,\cdot)$ in  the form 
\begin{equation}\label{optimaljoint} q_{0T}^\circ(x_0,x_T)=\hat{\varphi}(x_0) p(0,x_0;T,x_T)\varphi(x_T),
\end{equation} where $\varphi$ and $\hat{\varphi}$ {must satisfy}
\begin{eqnarray}\hat{\varphi}(x_0)\sum_{x_T}p(0,x_0;T,x_T)\varphi(x_T)&=&\mathbf p_0(x_0),\label{opt1}\\
\varphi(x_T)\sum_{x_0}p(0,x_0;T,x_T)\hat{\varphi}(x_0)&=&\mathbf p_T(x_T).\label{opt2}
\end{eqnarray}
Let us define $\hat{\varphi}(0,x_0)=\hat{\varphi}(x_0)$, $\quad \varphi(T,x_T)=\varphi(x_T)$ and  
$$\hat{\varphi}(T,x_T)=\sum_{x_0}p(0,x_0;T,x_T)\hat{\varphi}(0,x_0),\quad \varphi (0,x_0):=\sum_{x_T}p(0,x_0;T,x_T)\varphi(T,x_T).
$$
Then, (\ref{opt1})-(\ref{opt2}) can be replaced by the system
\begin{eqnarray}\label{Schonestep1}
\hat{\varphi}(T,x_T)=\sum_{x_0}p(0,x_0;T,x_T)\hat{\varphi}(0,x_0),\\\label{Schonestep2}\quad \varphi (0,x_0):=\sum_{x_T}p(0,x_0;T,x_T)\varphi(T,x_T)
\end{eqnarray}
with the boundary conditions
\begin{equation}\label{BConestep}
\varphi(0,x_0)\cdot\hat{\varphi}(0,x_0)=\mathbf p_0(x_0),\quad \varphi(T,x_T)\cdot\hat{\varphi}(T,x_T)=\mathbf p_T(x_T),\quad \forall x_0, x_T\in\mathcal X.
\end{equation}
The question of existence and uniqueness of  functions $\hat{\varphi}(x_0)$, $\varphi(x_T)$ satisfying (\ref{Schonestep1})-(\ref{Schonestep2})-(\ref{BConestep}) will be established in Section \ref{solonestep}. Before we do that, however, we investigate what else can be said about the solution when the prior random vector happens to be a ``time window" of a Markov chain.

\subsection{Markovian prior}\label{sec:prior}
Consider the special case where $X=\{X(t), 0\le t\le T\}$ is a window of a Markov chain. Let us introduce, in the language of Doob, the {\em space-time harmonic} function
\begin{equation}\label{defvarphi}\varphi (t,x_t):=\sum_{x_T}p(t,x_t;T,x_T)\varphi(x_T),\quad 0\le t\le T,
\end{equation}
and the {\em space-time co-harmonic} function 
\begin{equation}\label{defvarphihat}\hat{\varphi} (t,x_t):=\sum_{x_0}p(0,x_0;t,x_t)\hat{\varphi}(x_0), \quad 0\le t\le T.
\end{equation}
Because of the Markov property, for $0\le s<t<u\le T$, we have
$$
p(s,x_s;u,x_u)=\sum_{x_t}p(s,x_s;t,x_t)p(t,x_t;u,x_u).
$$
For compactness, we often write $\pi_{x_t,x_{t+1}}(t)$  and $\pi_{x_t,x_{t+n}}^{(n)}(t)$ of instead of $p(t,x_t;t+1,x_{t+1})$ and $p(t,x_t;t+n,x_{t+n})$, respectively. Hence, $\varphi$ and $\hat{\varphi}$ satisfy the backward and forward equation, respectively,
\begin{eqnarray}
&&\varphi(t,x_t)=\sum_{x_{t+1}}\pi_{x_t,x_{t+1}}(t)\varphi(t+1,x_{t+1}),\\&&\hat{\varphi}(t+1,x_{t+1})=\sum_{x_t}\pi_{x_t,x_{t+1}}(t)\hat{\varphi}(t,x_t).
\end{eqnarray}
Let $q^\circ_t$ denote the distribution of the Schr\"{o}dinger bridge at time $t$. We get
\begin{eqnarray}
\nonumber q^\circ_t(x_t)=\sum_{x_0}\sum_{x_T}p(0,x_0;t,x_t;T,x_T)q_{0T}^\circ(x_0,x_T)=\\\sum_{x_0}\sum_{x_T}\frac{p(0,x_0;t,x_t)p(t,x_t;T,x_T)}{p(0,x_0;T,x_T)}\hat{\varphi}(x_0)p(0,x_0;T,x_T)\varphi(x_T)\nonumber\\=\hat{\varphi} (t,x_t)\cdot\varphi (t,x_t).\label{factor}
\end{eqnarray}
Similarly, one gets
$$q_{st}^\circ(x_s,x_t)=\sum_{x_0}\sum_{x_T}\hat{\varphi}(x_0)p(0,x_0;s,x_s)p(s,x_s;t,x_t)p(t,x_t;T,x_T)\varphi(x_T)
$$
which yields, using (\ref{defvarphi})-(\ref{defvarphihat}) and (\ref{factor}), the new transition probabilities
\begin{equation}\label{TRA}q^\circ(s,x_s;t,x_t)=\frac{q_{st}^\circ(x_s,x_t)}{q^\circ_s(x_s)}=p(s,x_s;t,x_t)\frac{\varphi(t,x_t)}{\varphi(s,x_s)}.
\end{equation}
Notice, in particular, that the Schr\"{o}dinger bridge is also a {\em Markov chain} which is obtained from  the {\em a priori} model via a suitable {\em multiplicative functional transformation} just like in the diffusion case \cite{jamison1975Markovprocesses}. We have therefore established the following result \cite[Theorem 4.1]{pavon2010discrete}:
\begin{thm} \label{fund theorem}Assume that $p(0,\cdot; T,\cdot)$ is everywhere positive on $\mathcal X\times \mathcal X$. Suppose there exist positive functions  $\varphi$ and $\hat{\varphi}$ defined on $[0,T]\times\mathcal X$ satisfying for $t\in[0,T-1]$ the system
\begin{eqnarray}\label{Schroedingersystem1}
\varphi(t,x_t)&=&\sum_{x_{t+1}}\pi_{x_t,x_{t+1}}(t)\varphi(t+1,x_{t+1}),\\\hat{\varphi}(t+1,x_{t+1})&=&\sum_{x_t}\pi_{x_t,x_{t+1}}(t)\hat{\varphi}(t,x_t).\label{Schroedingersystem2}
\end{eqnarray}
with the boundary conditions 
\begin{equation}\label{bndconditions}
\varphi(0,x_0)\cdot\hat{\varphi}(0,x_0)=\mathbf p_0(x_0),\quad \varphi(T,x_T)\cdot\hat{\varphi}(T,x_T)=\mathbf p_T(x_T),\quad \forall x_0, x_T\in\mathcal X.
\end{equation}
Then the Markov distribution $Q^\circ$ in ${\mathcal P}(\mathbf p_0,\mathbf p_T)$ with one-step transition probabilities
\begin{equation}\label{opttransition}\pi_{x_t,x_{t+1}}^\circ(t)=\pi_{x_t,x_{t+1}}(t)\frac{\varphi(t+1,x_{t+1})}{\varphi(t,x_t)}
\end{equation}
is the unique solution of Problem \ref{prob:optimization}.
\end{thm}

Notice that, if a pair $\varphi,\hat{\varphi}$ solves (\ref{Schroedingersystem1})-(\ref{Schroedingersystem2})-(\ref{bndconditions}), so does the pair $a\varphi,a^{-1}\hat{\varphi}$ for $a>0$. This arbitrariness in the scaling, however, does not affect the transition probabilities (\ref{opttransition}). 

Schr\"{o}dinger and Kolmogorov were struck by the intrinsic time-reversibility of the solution: Swapping $\mathbf p_0$ and $\mathbf p_T$ leads to a solution bridge which is simply the time reversal of the original one. Moreover, the factorization (\ref{factor}) resembles Born's relation between the probability density and the wave function in quantum mechanics $\rho_t(x)=\psi(x,t)\psi^\dagger(x,t)$ where $\dagger$, throughout the paper, denotes conjugation/adjoint in a complex Hilbert space \footnote{Schr\"{o}dinger:``Merkw\"{u}rdige Analogien zur
Quantenmechanik, die mir sehr  des Hindenkens wert erscheinen" (remarkable analogies to quantum mechanics which appear to me very worth of reflection). Recall that Schr\"{o}dinger never accepted the so-called orthodox theory of measurement in quantum mechanics and was looking for a more classical probabilistic reformulation.}.

At this point, it should be apparent that the bottleneck of the theory of Schr\"{o}dinger bridges is the existence and uniqueness (up to multiplication by a positive constant) of the pair $\varphi,\hat{\varphi}$ solving the Schr\"{o}dinger system (\ref{Schonestep1})-(\ref{Schonestep2})-(\ref{BConestep}) (in the Markov case,  (\ref{Schroedingersystem1})-(\ref{Schroedingersystem2})-(\ref{bndconditions})). Schr\"{o}dinger  thought that existence and uniqueness should  be guaranteed in the diffusion case  ``except possibly for very nasty $\rho_0$, $\rho_T$, since the question leading to the pair of equations is so reasonable".  This problem turns out to be quite nontrivial and was settled in various degrees of generality by  Beurling, Jamison and F\"{o}llmer  \cite{beurling1960automorphism,jamison1974reciprocal,foellmer1988randomfields} establishing the feasibility of the dual optimization problem of Problem \ref{prob:optimization}. There is, however, an alternative approach based on proving convergence of successive approximations by Fortet \cite {fortet1940}. As it turns out, this more algorithmic approach had independent counterparts called ``iterative fitting algorithms" in the statistical literature on contingency tables \cite{demingstephan1940}. These were later shown to converge to a ``minimum discrimination information" \cite{irelandkullback1968,fienberg1970}, namely to a minimum entropy distance, see also \cite {csiszar1975divergence}. It will be apparent below that the approach pioneered by Fortet features a very desirable property: While establishing existence and uniqueness for the Schr\"{o}dinger system, it also provides a computationlly efficient  {\em algorithm} to actually {\em compute} the space-time harmonic function $\varphi$ and therefore the solution.

Our approach consists in showing that a certain iterative scheme for the Schr\"{o}dinger system is a contraction mapping on the positive orthant with respect to the projective metric which was introduced in Section \ref{sec:hilbert}. We argue that this is the {\em natural} metric in which to cast the iteration for these problems both in the classical and in the quantum case. Before we turn to that, however, we show that in the Markov case the new transition mechanism may be obtained via the solution of a two consecutive times interval problem.

\subsection{Reduction to the one-step bridge problem}\label{reduction}

Let $\Pi(t)=\left(\pi_{x_t,x_{t+1}}(t)\right)_{x_t,x_{t+1}=1}^N$  and $\Pi^\circ(t)=\left(\pi^\circ_{x_t,x_{t+1}}(t)\right)_{x_t,x_{t+1}=1}^N$ be the transition matrices of the prior distribution and of the bridge distribution. It is interesting to express (\ref{opttransition}) in matrix form. To this end, we introduce the notation
\[
\phi(t) = \diag\left(\varphi(t,x_1),\varphi(t,x_2),\ldots,\varphi(t,x_N)\right)
\]
for a diagonal matrix formed out of the entries of $\varphi(t,\cdot)$.
We now have
\begin{equation}\label{opttransitionmatrix}
\Pi^\circ(t)=\phi(t)^{-1}\Pi(t)\phi(t+1).
\end{equation}
Consider now  for $n\ge 1$ $\Pi^{(n)}(t)=\left(\pi^{(n)}_{x_t,x_{t+n}}(t)\right)_{x_t,x_{t+n}=1}^N$  the $n$-step transition probabilities matrix. By the Markov property, we have 
$$\Pi^{(n)}(t)=\Pi(t)\cdot\Pi(t+1)\cdots\Pi(t+n-1).
$$
Being the product of stochastic matrices, $\Pi^{(n)}(t)$ is also {\em stochastic}\footnote{The elements of $\Pi^{(n)}(t)$ are nonnegative as sum of products of nonnegative numbers. Moreover, let $\mathds{1}^\dagger=(1,1,\ldots,1)$. Then the fact that a matrix $Q$ has rows summing to one can be expressed as $Q \mathds{1}=\mathds{1}$.
As each $\Pi(t)$ has rows summing to one , we have
$$\Pi^{(n)}(t)\mathds{1}=\Pi(t)\cdot\Pi(t+1)\cdots\Pi(t+n-1)\mathds{1}
                                     =\Pi(t)\cdot\Pi(t+1)\cdots\Pi(t+n-2)\mathds{1}
                                     =\cdots=\Pi(t)\mathds{1}
                                     =\mathds{1}.
$$}.
Consider now $\Pi^{\circ^{(n)}}(t)$. By (\ref{opttransitionmatrix}), we get
\begin{eqnarray}\nonumber
&&\Pi^{\circ^{(n)}}(t)=\Pi^\circ(t)\cdot \Pi^\circ(t+1)\cdots \Pi^\circ(t+n-1)\\&&=\phi(t)^{-1}\Pi^{(n)}(t)\phi(t+n).\nonumber
\end{eqnarray}
Thus, we get the following remarkable generalization of formula (\ref{opttransition})
\begin{equation}\label{opttransitionnstep}
\pi_{x_tx_{t+n}}^{\circ^{(n)}}(t)=\pi^{(n)}_{x_t,x_{t+n}}(t)\frac{\varphi(t+n,x_{t+n})}{\varphi(t,x_t)}.
\end{equation}
Consider now $\Pi^{(T)}(0)$ as the transition matrix of a prior ``stroboscopic" evolution. Consider also the Schr\"{o}dinger bridge problem with the same marginals $\mathbf p_0$ and $\mathbf p_T$ as before at the two {\em consecutive} times $0$ and $T$. Specializing Theorem \ref{fund theorem} to this simple situation we get that the new transition probabilities are precisely given by (\ref{opttransitionnstep}) with $n=T$ where $\varphi$, $\hat{\varphi}$ satisfy (\ref{Schonestep1})-(\ref{Schonestep2})-(\ref{BConestep}), namely 
\begin{eqnarray}\label{Schroedingersystemonestep}
\varphi(0,x_0)&=&\sum_{x_T} \pi^{(T)}_{x_0,x_T}(0)\varphi(T,x_T),\\\hat{\varphi}(T,x_T)&=&\sum_i\pi^{(T)}_{x_0,x_T}(0)\hat{\varphi}(0,x_0).
\end{eqnarray}
with the boundary conditions 
\begin{equation}\label{bndconditionsonestep}
\varphi(0,x_0)\cdot\hat{\varphi}(0,x_0)=\mathbf p_0(x_0),\quad \varphi(T,x_T)\cdot\hat{\varphi}(T,x_T)=\mathbf p_T(x_T),\quad \forall x_0, x_T\in\mathcal X.
\end{equation}
We see that the solution of Problem \ref{prob:optimization} yields, as a by-product, the solution of the one-step problem. The converse, however, is also true. Solving (\ref{Schroedingersystemonestep})-(\ref{bndconditionsonestep}), yields the correct terminal value $\varphi(T,\cdot)$ from which $\varphi$ can be computed at all times through the iteration (\ref{Schroedingersystem1}). From it, the new transition probabilities are obtained through (\ref{opttransition}). We already know from Section \ref {discreterandom} that
solving  the one-step problem suffices to characterize the optimal distribution $Q^\circ$.  The argument above shows that it also permits to obtain the more explicit description, namely the transition mechanism of $Q^\circ$, which is desirable in the Markov case.

\subsection{The solution to the one-step bridge problem}\label{solonestep}

Motivated by what we have seen in Sections \ref{discreterandom} and \ref{reduction}, we study the one-step bridge problem with possibly non Markovian prior. The following result establishes existence and uniqueness for the system (\ref{Schonestep1})-(\ref{Schonestep2})-(\ref{BConestep}).
\begin{thm} \label{fundtheorem} Given a $N\times N$ stochastic matrix
\[
\Pi=\left[\pi_{x_0,x_T}\right]_{x_0,x_T=1}^N
\]
with strictly positive entries {\rm (}$\pi_{x_0,x_T}>0${\rm )} and probability distributions ${\mathbf p}_0$, ${\mathbf p}_T$, there exist four vectors
$
\varphi(0,x_0),\,\varphi(T,x_T),\,\hat\varphi(0,x_0),\,\hat\varphi(T,x_T)$, indexed by $x_0, x_T\in\mathcal X,
$
with positive entries such that
\begin{subequations}\label{eq:iteration}
\begin{eqnarray}\label{eq:iterationa}
\varphi(0,x_0)&=&\sum_{x_T}\pi_{x_0,x_T}\varphi(T,x_T),\\
\hat\varphi(T,x_T)&=&\sum_{x_0}\pi_{x_0,x_T}\hat\varphi(0,x_0),\label{eq:iterationb}\\
\label{eq:iterationc}
\varphi(0,x_0)\hat\varphi(0,x_0)&=&{\mathbf p}_0(x_0),\\\label{eq:iterationd}
\varphi(T,x_T)\hat\varphi(T,x_T)&=&{\mathbf p}_T(x_T).
\end{eqnarray}
\end{subequations}
The four vectors are unique up to multiplication of $\varphi(0,x_0)$ and $\varphi(T,x_T)$ by the same positive constant and division of $\hat\varphi(0,x_0)$ and $\hat\varphi(T,x_T)$ by the same constant.
\end{thm}
Rather than relying on the results of Beurling and Jamison as in \cite{pavon2010discrete,ticozzi2010time}, we give an independent proof which also yields an effective algorithm. The proof relies on showing that a certain iteration is strictly contractive in the Hilbert metric\footnote{Fortet's proof, which we find very difficult to follow, is apparently based on establishing monotonicity of two sequences of functions produced by the iteration.}.
\begin{lemma}\label{keylemma} Consider the following circular diagram of maps
\begin{equation}
\begin{array}{ccccc}
&\hat\varphi(0,x_0) & \overset{\cE^\dagger}{\longrightarrow} & \hat\varphi(T,x_T) & =\sum_{x_0}\pi_{x_0,x_T}\hat \varphi(0,x_0)
\\\\
\hat\varphi(0,x_0)= \frac{{\mathbf p}_0(x_0)}{\varphi(0,x_0)}& \uparrow & & \downarrow &\varphi(T,x_T)= \frac{{\mathbf p}_T(x_T)}{\hat\varphi(T,x_T)}\\\\
\sum_{x_N}\pi_{x_0,x_N}\varphi(T,x_T)=&\varphi(0,x_0) &\overset{\cE}{\longleftarrow} & \varphi(T,x_T) &
\end{array}
\end{equation}
where
\begin{eqnarray*}
{\hat \cD}_0\;:\;\varphi(0,x_0)&\mapsto& \hat\varphi(0,x_0)= \frac{{\mathbf p}_0(x_0)}{\varphi(0,x_0)}\\
\cD_T\;:\;\hat\varphi(T,x_T)&\mapsto& \varphi(T,x_T)=\frac{{\mathbf p}_T(x_N)}{\hat\varphi(T,x_T)}
\end{eqnarray*}
represent componentwise division of vectors.
Then the composition
\begin{equation}\label{eq:composition}
\hat\varphi(0,x_0)  \overset{\cE^\dagger}{\longrightarrow} \hat\varphi(T,x_T)\overset{\cD_{T}}{\longrightarrow}\varphi(T,x_T)
\overset{\cE}{\longrightarrow} \varphi(0,x_0)
\overset{\cD_{0}}{\longrightarrow}\left(\hat\varphi(0,x_0)\right)_{\rm next}
\end{equation}
is contractive in the Hilbert metric. 
\end{lemma}
Before proceeding with the proof
we provide a note about notation: The map
\[
\cE^\dagger \;:\; \hat\varphi(0,x_0) \mapsto \hat\varphi(T,x_T)=\sum_{x_0}\pi_{x_0,x_N}\hat \varphi(0,x_0)
\]
is the adjoint
 of the backward evolution
\[
\cE\;:\; \varphi(T,x_T)\mapsto \varphi(0,x_0)=\sum_{x_N}\pi_{x_0,x_T}\varphi(T,x_T),
\]
which is consistent with the standard notation in diffusion processes where the Fokker-Planck (forward) equation involves the adjoint of the {\em generator} appearing in the backward Kolmogorov equation.

Also notice that the componentwise divisions of ${\hat \cD}_0$ and $\cD_T$ are well defined. Indeed, even when $\hat\varphi(0)$ ($\varphi(T)$) has zero entries, $\hat\varphi(T)$ ($\varphi(0)$) has all positive entries since the elements of $\Pi$ are all positive.

\noindent{\bf Proof of Lemma \ref{keylemma}:} 
The diameter of the range of $\cE$ is
\begin{eqnarray*}
\Delta(\cE)&=&\sup\{d_H(\cE(x),\cE(y)) \mid x_i>0,\,y_i>0\}\\
&=&\sup\{\log\left(\frac{\pi_{ij}\pi_{k,\ell}}{\pi_{i,\ell}\pi_{k,j}}\right)\mid1\leq  i,j,k,\ell\leq n\}
\end{eqnarray*}
is finite since all entries $\pi_{i,j}$'s are positive. 
Birkhoff's theorem {(Theorem \ref{BBcontraction})} provides a contraction coefficient for linear positive maps and, in this case,
{we have}
\[
\|\cE\|_H=\tanh(\frac{1}{4}\Delta(\cE))<1.
\]
For the adjoint map $\cE^\dagger$ we only need to note that it is homogeneous of degree $1$ and therefore, by Birkhoff's theorem,
\[
\|\cE^\dagger\|_H\leq 1.
\]
{Next we note that provided 
${\mathbf p}_0(x_0)$ and ${\mathbf p}_T(x_N)$ have positive entries, both ${\hat \cD}_0$ and $\cD_T$
are isometries in the Hilbert metric since inversion and element-wise scaling are both isometries. Indeed, for vectors $[x_i]_{i=1}^N$,  $[y_i]_{i=1}^N$, it holds that
\begin{eqnarray*}
d_H([x_i],[y_i])&=&\log\left((\max_i (x_i/y_i))\frac{1}{\min_i (x_i/y_i)}\right)\\
&=&\log\left(\frac{1}{\min_i ((x_i)^{-1}/(y_i)^{-1})}\max_i ((x_i)^{-1}/(y_i)^{-1})\right)\\&=&d_H([(x_i)^{-1}],[(y_i)^{-1}])
\end{eqnarray*}
and
\begin{eqnarray*}
d_H([{\mathbf p}_i x_i],[{\mathbf p}_i y_i])&=&\log\frac{\max_i (({\mathbf p}_i x_i)/({\mathbf p}_i y_i))}{\min_i (({\mathbf p}_i x_i)/({\mathbf p}_i y_i))}\\
&=&\log\frac{\max_i (x_i/y_i)}{\min_i (x_i/y_i)}=d_H([x_i],[y_i]).
\end{eqnarray*}
If ${\mathbf p}_0(x_0)$ and ${\mathbf p}_T(x_N)$ have any zero entries, then
${\hat \cD}_0$ and $\cD_T$
are in fact contractions.}
Finally, we observe that
\[
\|{\hat \cD}_0\circ \cE\circ \cD_T\circ\cE^\dagger\|_H\leq \|{\hat \cD}_0\|_H\cdot\|\cE\|_H\cdot\|\cD_T\|_H\cdot\|\cE^\dagger\|_H<1,
\]
where $\circ$ denotes composition. Therefore, the composition $\cC:={\hat \cD}_0\circ \cE\circ \cD_T\circ\cE^\dagger$ is contractive:  Namely, 
\[
d_H(\cC(x),\cC(y))\le\|\cC\|_Hd_H(x,y), \quad 0\le\|\cC\|_H<1 .
\]
This completes the proof of the lemma.\hfill $\Box$

\noindent{\bf Proof of Theorem \ref{fundtheorem}:}
{Since $\cC$ is contractive in the Hilbert metric, there is a unique positive $\hat\varphi(0,\cdot)=[\hat\varphi(0,x_0)]$ so that the corresponding ray is invariant under $\cC$. That is, in the notation of \eqref{eq:composition},
\begin{eqnarray*}
(\hat\varphi(0,\cdot))_{\rm next}&=&\cC(\hat\varphi(0,\cdot))\\
&=& \lambda \hat\varphi(0,\cdot),
\end{eqnarray*}
for the composition $\cC:={\hat \cD}_0\circ \cE\circ \cD_T\circ\cE^\dagger$. From this we can obtain
\begin{eqnarray*}
\hat\varphi(T,\cdot)&=&\cE^\dagger(\hat\varphi(0,\cdot)),\\
\varphi(0,\cdot)&=&\cE(\varphi(T,\cdot)),
\end{eqnarray*}
while
\begin{eqnarray*}
\hat\varphi(T,\cdot)\varphi(T,\cdot)&=& {\mathbf p}_T(\cdot)\mbox{ and}\\
\lambda\hat\varphi(0,\cdot)\varphi(0,\cdot)&=&{\mathbf p}_0(\cdot). 
\end{eqnarray*}
However, since
\begin{eqnarray*}
1&=& \lambda\langle \hat\varphi(0,\cdot),\varphi(0,\cdot)\rangle\\
&=& \lambda\langle \hat\varphi(0,\cdot),\cE(\varphi(T,\cdot))\rangle\\
&=&\lambda\langle \cE^\dagger(\hat\varphi(0,\cdot)),\varphi(T,\cdot)\rangle\\
&=&\lambda\langle \hat\varphi(T,\cdot),\varphi(T,\cdot)\rangle\\
&=&\lambda,
\end{eqnarray*}
we conclude that 
these satisfy the Schr\"{o}dinger system (\ref{eq:iterationa})-(\ref{eq:iterationb})-(\ref{eq:iterationc})-(\ref{eq:iterationd}).} \hfill $\Box$

\begin{remark} {\bf Numerical algorithm:} The step above suggest that starting from any positive vector, e.g., $x(0)=\mathds{1}$, the unique fixed point is given by
\begin{eqnarray}\label{eq:algorithm}
\hat\varphi(0) &=& \lim_{k\to\infty} x(k),
\end{eqnarray}
where $x(k+1)=\cC(x(k))$ for $k=1,2,\ldots$. 
\end{remark}
By the variational analysis of Section \ref{discreterandom}, we finally get the following result.
\begin{thm} Assume that $\mathbf p_0$ and $\mathbf p_T$ are positive on $\mathcal X$ and that $p(0,\cdot; T,\cdot)$ is everywhere positive on $\mathcal X\times \mathcal X$. Then, the unique solution to Problem \ref{prob:optimization} with prior $P\in\mathcal P$ is given by 
$$Q^\circ(x_0,x_1,\ldots,x_T)=P_{x_0,x_T}(x_0,x_1,\ldots,x_T)q^\circ_{0T}(x_0,x_T),$$
where $q^\circ_{0T}(x_0,x_T)$ solving the problem for the two consecutive times $0,T$ is
given by
\begin{subequations}\label{eq:solutiononestep}
\begin{equation}
q_{0T}^\circ(x_0,x_T)=\hat{\varphi}(0,x_0) p(0,x_0;T,x_T)\varphi(T,x_T).
\end{equation}
Here $\hat{\varphi}(0,x_0)$, $\varphi(T,x_T)$  are as in Theorem \ref{fundtheorem} with 
$$\Pi=\left[\pi_{x_0,x_T}\right]_{x_0,x_T=1}^N, \quad \pi_{x_0,x_T}=p(0,x_0;T,x_T).$$
\end{subequations}
\end{thm}

{The corresponding transition probability is given
by
\begin{equation}\label{eq:transition_q0T}
q^\circ(0,x_0;T,x_T)= p(0,x_0;T,x_T)\frac{\varphi(T,x_T)}{\varphi(0,x_0)}.
\end{equation}}\\
In fact, when $P$ is Markovian, we also get the new one-step transition probabilities in a similar manner, i.e., in the way described at the end of Section \ref{reduction}. {Finally, as already observed, extending the results of this section to the case of a countable state space is straightforward.}

\section{The Schr\"{o}dinger bridge problem for quantum evolutions}\label{sec:quantum}

We begin with some background on quantum probability and stochastic maps, {see e.g. \cite{nielsen-chuang2000quantum,holevo2001statistical,petz2008quantum}.}
In quantum probability there is neither a notion of a random variable nor of a probability space. Instead,
random variables are replaced by Hermitian matrices/operators, referred to as {\em observables}, while the expected outcome of an experiment is quantified by a suitable functional that provides the {\em expectation}. 
Throughout, we restrict our attention to the case {of  finite-dimensional quantum systems with associated Hilbert space
isomorphic to $\mC^n$. Here} experiments take values from a finite alphabet and, thereby, the mathematical framework relies on algebras of finite matrices. In this case, the state of the underlying system which defines the expectation is represented by a non-negative definite matrix with trace one which is referred to as a {\em density matrix}.
Notation and basic terminology are explained next, followed by the formulation the quantum bridge problem, namely, the problem of identifying a quantum channel that is consistent with given quantum states at its two ends and has a specific relation to a given prior quantum channel.
For potential applications of quantum bridges and links to topics in quantum information we refer to \cite[Section VII]{pavon2010discrete}.

\subsection{Quantum channels}\label{sec:quantumchannels}

We use 
$\fM:=\{X\in \mC^{n\times n} \}$ to denote square matrices, $\fH:=\{X\in \mC^{n\times n} \mid X=X^\dagger\}$ to denote the set of Hermitian matrices, $\fH_+$ and $\fH_{++}$ the cones of nonnegative and positive definite ones, respectively, and $\fD:=\{\rho\in\fH_+\mid \trace \rho=1\}$ the set of density matrices. The latter represents possible ``states'' of a quantum system. In turn, a state $\rho$ defines an {\em expectation} functional on observables $X\in\fH$
\[
\mE_\rho(X)=\trace(\rho X).
\]

The standard model for a quantum experiment as well as for a quantum channel is provided by a linear {\em trace-preserving completely-positive} (TPCP) map between density matrices (which may possibly be of different size). This is the quantum counterpart of a Markov evolution and is referred to as a {\em Kraus map}. For the {finite-dimensional} case treated herein, where quantum systems are represented by density matrices and where, for notational simplicity, matrices are all of the same size, a Kraus map assumes the representation
\begin{subequations}\label{eq:krausrepresentation}
\begin{equation}\label{eq:kraus}
\cE^\dagger \;:\; \fD\to\fD\;:\;\rho \longrightarrow \sigma=\sum_{i=1}^{n_\cE} E_i\rho E_i^\dagger,
\end{equation}
with $E_i\in\fM$ such that
\begin{equation}\label{eq:I}
\sum_{i=1}^{n_\cE}E_i^\dagger E_i=I,
\end{equation}
\end{subequations}
see \cite[Chapter 2]{petz2008quantum}. Throughout, $I$ denotes the identity matrix.
Condition \eqref{eq:I} ensures that the map preserves the trace, i.e., that $\trace(\sigma)=\trace(\rho)=1$. {\em Completely-positive} refers to the property that, besides the fact that $\cE^\dagger(\rho)\geq 0$ for all $\rho\geq 0$ (i.e., being a {\em positive} map),  if $\bI_k$ denotes the identity map on $\mC^{k\times k}$ and $\otimes$ the tensor product, then $\bI_k\otimes \cE^\dagger$ is a positive map for all $k$. The Kraus-representation (\ref{eq:kraus}-\ref{eq:I}) characterizes completely positive maps (see, e.g., \cite[Chapter 2]{petz2008quantum}).

Herein, a further condition will be imposed on Kraus maps which we 
refer to as {\em positivity improving}. This is the property that
\begin{equation}\label{eq:strictpositivity}
\cE^\dagger (\rho) >0 \mbox{ for any }\rho\in\fD.
\end{equation}
Thus, positivity improving amounts to having the range of $\cE^\dagger$ contained in the  interior of $\fD$ or, in other words and in view of \eqref{eq:kraus}, that there is no pair of vectors $w,\,v\in\mC^n$ such that $w^\dagger E_i v=0$ for all $i\in\{1,\ldots,n_\cE\}$. 
A necessary condition\footnote{To see this assume that $n_{\cE}\leq n$ and take $v$ to be an eigenvector of the matrix pencil $E_1-\lambda E_2$. Then $E_1v$ and $E_2v$ are linearly dependent and the  span of $E_1v$ \ldots $E_nv$ has dimension less than $n$. This implies that there exists a $w$ such that $w^\dagger E_iv=0$ for all $i$. Determining whether a given Kraus map actually has the property of positivity improving, in general, is NP-hard \cite{gaubert2014checking,hillar2009most}
}
for positivity improving is that $n_{\cE}> n$.

Our notation in using $\cE^\dagger$ for the Kraus map is \underline{not}  \underline{standard} and differs from that in e.g., \cite{petz2008quantum} where $\cE$ is used instead. However, our choice maintains consistency with the standard convention throughout probability theory {where the (forward) Fokker-Planck equation involves} the adjoint of the generator of a Markov semigroup.

A model for quantum measurement amounts to an instantiation of a quantum channel \cite[Chapter 2]{petz2008quantum}. Typically an observable $X\in\fH$ is specified having a decomposition
\begin{equation}\nonumber
X=\sum_i \lambda_i \x_i^\dagger \x_i 
\end{equation}
with $\x_i\in\fM$ such that
\[
\sum_i \x_i^\dagger \x_i=I.
\]
A particular case is that of a spectral decomposition of $X$. When the outcome $i$ is being recorded and the value $\lambda_i$ registered for a quantum system that is initially in a state $\rho$, then, after the measurement, the system finds itself in the new state
\begin{equation}\label{eq:averaging}
\left(\frac{1}{\trace (\x_i\rho \x^\dagger_i)}\right) \x_i\rho \x^\dagger_i.
\end{equation}
The likelihood of this particular outcome associated with recording $\lambda_i$ is $\trace (\x_i\rho \x^\dagger_i)$. Therefore,
\begin{align*}
\mE_\rho(X)&=\sum_i \lambda_i\trace (\x_i\rho \x^\dagger_i)\\
&=\trace(\rho X)
\end{align*}
in agreement with the earlier statement. On the other hand, if the experiment takes place but no particular outcome is recorded or, equivalently, all possible states in \eqref{eq:averaging} are weighed in by the corresponding probabilities $\trace (\x_i\rho \x^\dagger_i)$, the new state becomes
\[
\sum_i  \x_i \rho \x_i^\dagger.
\]
This is precisely a Kraus map acting on $\rho$ and this type of measurement is referred to as  {\em nonselective}.
In the sequel we will encounter observables $\phi\in\fH$ 
factored as $\phi=\x^\dagger \x$. These represent quantum analogues of the random variables $\varphi$ in the classical case. The expression
$\x \rho \x^\dagger$
may be thought of as an ``unnormalized'' state that results in after a nonzero outcome has been recorded when a measurement performed. The factors of observables $\phi$ will give rise to the multiplicative {functional} transformations on Kraus maps sought in addressing the quantum bridge problem. These specify Kraus maps that link a set of two given {``marginal''} density matrices.

While there are many similarities between quantum channels and Markov evolution, there are also stark differences. A fundamental departure from classical probability arises in that, in general, there is no notion of joint probablitity between ``measurements'' at the two ends of a quantum channel. In this, the order with which measurements take place matters.


\subsection{The bridge problem}\label{sec:quantumbridge}

Consider a reference quantum evolution given by an initial density matrix $\sigma_0$ and by a sequence of TPCP maps $\{\cE^\dagger_t; 0\le t\le T-1\}$, each admitting a Kraus representation with matrices $\{E_{t,i}\}$, so that
\[
\cE^\dagger_t : \sigma_t \mapsto \sigma_{t+1}=\sum_i E_{t,i} \sigma_t E_{t,i}^\dagger, \quad t=0,1,\ldots, T-1.
\]
As usual, 
\[
\sum_i E_{t,i}^\dagger E_{t,i} = I.
\] 
Consider also the composition map
\[
\cE_{0:T}^\dagger:=\cE^\dagger_{T-1}\circ\cdots \circ\cE^\dagger_1\circ\cE^\dagger_0.
\]
Observe that $\cE_{0:T}^\dagger$ is a TPCP map. For instance, it is immediate that $\cE^\dagger_1\circ\cE^\dagger_0$ admits a (double-indexed) Kraus representation with matrices $\{E_{1,j}E_{0,i}\}$. Hence, as thoroughly argued in the classical case, we only need to consider the one-step situation. Besides the reference quantum evolution, we are also given an initial and a final positive definite density matrices $\rho_0$ and $\rho_T$, respectively.

We consider the problem of finding a new quantum evolution $\cF_{0:T}^\dagger=\cF^\dagger_{T-1}\circ\cdots \circ\cF^\dagger_1\circ\cF^\dagger_0$, where the $\cF^\dagger_t$ are TPCP maps, such that the new evolution is close to the reference one and
\begin{equation}\label{eq:matchmarginals}
\cF_{0:T}^\dagger(\rho_0)=\rho_T.
\end{equation}
Providing a mathematical formulation for the selection of $\cF_t$'s is nontrivial. In \cite{pavon2010discrete}, which deals with the simpler problem where only the initial or final density matrix is prescribed,  {\em quantum paths} were introduced via a family of observables $\{X_t; 0\le t\le T\}$ with spectral decomposition
\[X_t=\sum_{i_t=1}^{m_t}x_{i_t}\Pi_{i_t}(t).
\]
Then, paths are defined as sequences of ``events'' $\left(\Pi_{i_0}(0),\Pi_{i_1}(1),\ldots,\Pi_{i_T}(T)\right)$ and thereby, one can define path-conditioned density evolutions and formulate and solve maximum entropy problems in a somewhat classical-like fashion. However, this approach does not seem to work when both initial and final densities are given, specifically because it is not clear what notion of relative entropy one should use to compare the two Markovian evolutions associated to $\cE$ and $\cF$. 
Thus, herein we formulate the quantum Shr\"odinger bridge problem as that of seeking a suitable multiplicative functional transformation of the prior Kraus evolution so as to meet the marginal conditions. This is explained next.

Given the sequence of TPCP maps $\{\cE^\dagger_t, \mbox{ for }0\le t\le T-1\}$ and the two marginals $\rho_0,\rho_T$ we seek a sequence of invertible matrices $\x_0,\ldots,\x_T$ such that for $t\in\{0,\ldots,T-1\}$,
\begin{equation}\label{eq:multtrans}
\cF^\dagger_t(\cdot)=\x_{t+1}\left(\cE_t^\dagger(\x_t^{-1}(\cdot)\x_t^{-\dagger})\right)\x_{t+1}^\dagger
\end{equation}
is a Kraus map (i.e., totally positive and trace preserving) and the evolution
\begin{eqnarray}
\cF_{0:T}^\dagger(\cdot)&=&\cF^\dagger_{T-1}\circ\cdots \circ\cF^\dagger_1\circ\cF^\dagger_0(\cdot)\\&=&
\x_T\left(\cE_{0:T}^\dagger(\x_0^{-1}(\cdot)\x_0^{-\dagger})\right)\x_T^\dagger\label{eq:multiplicative}
\end{eqnarray}
is consistent with the given marginals (i.e., \eqref{eq:matchmarginals} holds).
Conditions \eqref{eq:multtrans} and \eqref{eq:multiplicative} represent the quantum analogue of a multiplicative functional transformation.

Very much as in the classical case, the solution of the Shr\"odinger bridge in the multi-step case reduces to solving the one-step bridge problem:
given the triple $(\cE_{0:T}^\dagger,\rho_0,\rho_T)$, determine invertible matrices $\x_0,\x_T$ such that  \eqref{eq:multiplicative} is Kraus map and satisfies
\begin{eqnarray}\label{eq:agreement}
\cF_{0:T}^\dagger(\rho_0)&=&\rho_T.
\end{eqnarray}
The {\em unital} property of $\cF$, namely the fact that $\cF_{0:T}(I)=I$, when expressed in terms of $\cE$ and the factors in \eqref{eq:multiplicative}, implies that
\begin{eqnarray}\label{st-harmonic}
\cE_{0:T}(\phi_T)&=&\phi_0
\end{eqnarray}
for the Hermitian, positive definite matrices
\begin{equation}\label{FACTOR}\phi_0 = \x_0^\dagger \x_0,\quad
\phi_T = \x_T^\dagger \x_T.
\end{equation}
Likewise, in view of \eqref{eq:multtrans} and the requirement that $\cF^\dagger_t$ is trace preserving, we set
\begin{equation}\label{FACTOR2}
\cE_t(\phi_{t+1})=\phi_t,
\end{equation}
for $t\in\{T-1,\ldots,0\}$ and factor
\begin{equation}
\phi_t = \x_t^\dagger \x_t.\label{eq:spacetime}
\end{equation}
Conditions (\ref{st-harmonic}) and \eqref{eq:spacetime} indicate that the $\phi_t$'s are {\em space-time harmonic} matrices with respect to $\cE_{0:T}^\dagger$ and $\cE_t^\dagger$, respectively. On the other hand,
\begin{eqnarray}\label{eq:agreement2}
\cE_{0:T}^\dagger(\underbrace{\x_0^{-1}(\rho_0)\x_0^{-\dagger}}_{\hat\phi_0})&=&\underbrace{\x_T^{-1}(\rho_T)\x_T^{-\dagger}}_{\hat\phi_T},
\end{eqnarray}
where the matrices $\hat\phi_t$ in \eqref{eq:agreement2} can be thought of as unnormalized density matrices in a dual Heisenberg picture.

The concept of space-time harmonic functions for quantum channels was introduced and studied  in \cite{pavon2010discrete,ticozzi2010time}.
Thus, following \cite{pavon2010discrete,ticozzi2010time}, when two Kraus maps are related as in (\ref{eq:multiplicative}-\ref{st-harmonic}-\ref{FACTOR}), we say that $\cF^\dagger$ is obtained from $\cE^\dagger$ through a {\em multiplicative functional transformation induced by a space-time harmonic function}. In  \cite{pavon2010discrete}, it was shown that the solution of the Schr\"{o}dinger problem where only the initial or final density is assigned is indeed given by a multiplicative functional transformation of the prior providing the first such example of a non-commutative counterpart of the classical results. A variational characterisation of the quantum bridge as a critical point of a suitable entropic functional for the present general setting is unknown.

\subsection{Doubly stochastic Kraus maps}\label{sec:doublystochastic}

The results we present next are for the special case where the two marginal densities are uniform, i.e., when both
\[
\rho_0=\frac{1}{n}I \mbox{ as well as }\rho_T=\frac{1}{n}I.
\]
This special case is interesting in its own right as, already for the classical case, it represents a well known basic result in the statistics literature (Sinkhorn's theorem) \cite{Sinkhorn1964,sinkhorn1974diagonal}. 
Sinkhorn's result states that for any stochastic matrix $[\pi_{x_0,x_T}]_{x_0,x_T=1}^N$ {\em having all entries strictly positive}, there exists a unique multiplicative {functional} transformation
\begin{eqnarray*}
\pio_{x_0,x_T}&=&\pi_{x_0,x_T}
\frac{\varphi(T,x_T)}{\varphi(0,x_0)}
\end{eqnarray*}
so that $\pio\;$ is doubly stochastic, that is, it has nonnegative elements and satisfies
\[
\sum_{x_0}\pio_{x_0,x_T}=\sum_{x_T}\pio_{x_0,x_T}=1.
\] 
{In view of \eqref{eq:transition_q0T}, in this classical setting, Sinkhorn's result is a direct corollary of Theorem
\ref{fundtheorem}.}

The property of a stochastic matrix to have all entries strictly positive corresponds to the positivity improving property \eqref{eq:strictpositivity} of a Kraus map. {We proceed to derive the quantum counterpart of Sinkhorn's result.}
\begin{thm}\label{thm:doublystochastic0}
Given a positivity improving Kraus map $\cE_{0:T}^\dagger$, i.e., satisfying \eqref{eq:strictpositivity}, there exists a pair of observables
$\phi_0,\,\phi_T\in\fH_{++}$ unique up to multiplication by a positive constant {and related as in (\ref{st-harmonic})} such that,
for any factorization
\begin{eqnarray*}
\phi_0 &=& \x_0^\dagger \x_0,\mbox{ and}\\ 
\phi_T &=& \x_T^\dagger \x_T,
\end{eqnarray*}
the map $\cF_{0:T}^\dagger:\fD\to\fD$ defined by
\begin{equation}\label{eq:doublystochastic2}
\cF^\dagger(\cdot):= \x_T\left(\cE_{0:T}^\dagger(\x_0^{-1}(\cdot)\x_0^{-\dagger})\right)\x_T^\dagger
\end{equation}
is a {\em doubly stochastic} Kraus map, in that $\cF(I)=I$ as well as $\cF^\dagger(I)=I$. 
\end{thm}

Theorem \ref{thm:doublystochastic0} follows immediately from the following result that we establish first.

\begin{thm}\label{thm:doublystochastic}
Given a Kraus map $\cE_{0:T}^\dagger$ satisfying \eqref{eq:strictpositivity}, there exist observables $\phi_0$, $\phi_T$ in $\fH_{++}$ {unique up to multiplication by a positive constant} 
such that
\begin{align*}
\cE_{0:T}(\phi_T) &= \phi_0\\
\cE_{0:T}^\dagger(\phi_0^{-1}) &= \phi_T^{-1}.
\end{align*}
\end{thm}

\noindent{\bf Proof of Theorem \ref{thm:doublystochastic}:} 
The claim in Theorem \ref{thm:doublystochastic} amounts to the existence of a unique fixed point for the following circular diagram of maps
\begin{equation}\label{eq:circular}
\begin{array}{ccccc}
&\hat\phi_0 & \overset{\cE^\dagger_{0,T}}{\longrightarrow} & \hat\phi_T & 
\\\\
\hat\phi_0= \phi_0^{-1}& \uparrow & & \downarrow &\phi_T= \hat\phi_T^{-1}\\\\
&\phi_0 &\overset{\cE_{0,T}}{\longleftarrow} & \phi_T &
\end{array}
\end{equation}
Thus, it suffices to show that the composition map
\begin{equation}\label{eq:contractivecomposition}
\cC\;:\;\left(\hat\phi_0\right)_{\rm starting}  \overset{\cE^\dagger_{0,T}}{\longrightarrow} \hat\phi_T\overset{(\cdot)^{-1}}{\longrightarrow}\phi_T
\overset{\cE_{0,T}}{\longrightarrow} \phi_0
\overset{(\cdot)^{-1}}{\longrightarrow}\left(\hat\phi_0\right)_{\rm next}
\end{equation}
from $\fH_{++}\to\fH_{++}$ is contractive in the Hilbert metric. It should be noted, that ``points'' here are defined up to a scaling factor, thus, they in essence represent rays.

Once again, we use Birkhoff's theorem to determine the contraction coefficient. Here, we are working on the positive cone $\fH_+$. This has a nonempty interior $\fH_{++}$ and the partial order defined by nonnegative definiteness. Accordingly,
\begin{eqnarray*}
M(X,Y)&=&\inf\{\lambda \mid X\leq \lambda Y\}\\
&=&\max\{{\rm eig}(Y^{-1/2}XY^{-1/2})\}\\
&=& \max\{{\rm eig}(XY^{-1})\},
\end{eqnarray*}
\begin{eqnarray*}
m(X,Y)&=&\sup\{\lambda \mid \lambda y\leq x\}\\
&=& \min\{{\rm eig}(XY^{-1})\},
\end{eqnarray*}
where ${\rm eig}(\cdot)$ denotes the ``eigenvalues of,'' and in this case
the Hilbert metric is
\begin{eqnarray*}
d_H(X,Y)&=&d_H(Y^{-1/2}XY^{-1/2},I)\\
&=&\log(\kappa(XY^{-1}))
\end{eqnarray*}
where $\kappa(\cdot)$ is the ``conditioning number'' of a $Z\in\fM$,
\begin{eqnarray*}
\kappa(Z)&=&\frac{\max\{{\rm eig}(Z)\}}{\min\{{\rm eig}(Z)\}}.
\end{eqnarray*}
From the Birkhoff-Bushell theorem, we have that both $\|\cE\|_H\leq 1$
as well as $\|\cE^\dagger\|_H\leq 1$ (from linearity together with monotonicity). Furthermore, 
the diameter of the range of $\cE$,
\begin{eqnarray*}
\Delta(\cE)&=&\sup\{d_H(\cE(X),\cE(Y)) \mid X,\,Y\in\fH_{++}\},
\end{eqnarray*}
is finite.
To see this first note that
\[
\Delta(\cE)\leq 2 \sup\{d_H(\cE(X),I) \mid X\in\fH_{++}\},
\]
utilizing the metric property and the fact that $\cE(I)=I$, and then
note that $d_H(\cE(X),I)$ is invariant under scaling of $X$ by a scalar. Therefore we can restrict our attention to $X\in\fD$ instead, and since $\cE(X)>0$ for all $X\in\fD$, by compactness of $\fD$,
\begin{eqnarray*}
\Delta(\cE)&\leq&2\max\{d_H(\cE(X),I) \mid X\in\fD\}<\infty.
\end{eqnarray*}
Therefore (see Theorem \ref{BBcontraction}), 
\[
\|\cE\|_H=\tanh(\frac{1}{4}\Delta(\cE))<1.
\]
Finally, we note that the induced Hilbert-gain of the inversion
\[
Z\mapsto Z^{-1}
\]
is $1$ since
\[
d_H(X,Y)=d_H(X^{-1},Y^{-1}).
\]
We conclude that with $\cC$ as in \eqref{eq:contractivecomposition}, $\|\cC\|_H<1$. {Again, 
contractiveness in the Hilbert metric, there exists a unique fixed ray, i.e.,
\[
\hat\phi_0=\lambda\cC(\hat\phi_0),
\]
for some $\lambda>0$, and corresponding
\begin{eqnarray*}
\hat\phi_T&=&\cE^\dagger_{0:T}(\hat\phi_0)\\
\phi_0&=&\cE_{0:T}(\phi_T)
\end{eqnarray*}
while
\begin{eqnarray*}
\phi_T&=&\hat\phi_T^{-1}\mbox{ and}\\
\lambda \hat\phi_0&=&\phi_0^{-1}.\\
\end{eqnarray*}
Since,
\begin{eqnarray*}
n&=&\lambda\trace(\hat\phi_0\phi_0)\\
&=&\lambda\trace(\hat\phi_0\cE_{0:T}(\phi_T))\\
&=&\lambda\trace(\cE_{0:T}^\dagger(\hat\phi_0)\phi_T)\\
&=&\lambda\trace(\hat\phi_T\phi_T)\\
&=&\lambda n,
\end{eqnarray*}
and therefore $\lambda=1$. This completes the proof of the theorem.}\hfill $\Box$

\noindent{\bf Proof of Theorem \ref{thm:doublystochastic0}:} 
This is truly a corollary to Theorem \ref{thm:doublystochastic}. Since $\phi_0,\,\phi_T$ in the proof of Theorem \ref{thm:doublystochastic} are in $\fH_{++}$, we take any factorization
\begin{eqnarray*}
\phi_0 &=& \x_0^\dagger \x_0,\mbox{ and}\\ 
\phi_T &=& \x_T^\dagger \x_T.
\end{eqnarray*}
Then, with $\cF^\dagger$ as in \eqref{eq:doublystochastic2},
\begin{eqnarray*}
\cF_{0:T}^\dagger(I)&=& \x_T\left[\cE_{0:T}^\dagger\left(\x_0^{-1}(I)\x_0^{-\dagger}\right)\right]\x_T^\dagger\\
&=&\x_T\left[\cE_{0:T}^\dagger(\phi_0^{-1})\right]\x_T^\dagger\\
&=&\x_T\left(\phi_T^{-1}\right)\x_T^\dagger\\
&=&I.
\end{eqnarray*}
Likewise, 
\begin{eqnarray*}
\cF_{0:T}(I)&=& \x_0^{-\dagger}\left[\cE_{0:T}\left(\x_T^{\dagger}(I)\x_T\right)\right]\x_0^{-1}\\
&=&\x_0^{-\dagger}\left[\cE_{0:T}(\phi_T)\right]\x_0^{-1}\\
&=&\x_0^{-\dagger}\left(\phi_0\right)\x_0^{-1}\\
&=&I.
\end{eqnarray*}
Thus, $\cF^\dagger$ is doubly stochastic as claimed.
\hfill $\Box$

While in the classical case (Sinkhorn's theorem) there is a unique doubly stochastic map obtained from $\pi_{x_0,x_T}$ via a multiplicative {functional} transformation, in the quantum case this is clearly false. Uniqueness in Theorem \ref{thm:doublystochastic0} is claimed for the observables $\phi_0,\phi_T$. Hence, $\cF$ is unique modulo corresponding unitary factors; obviously, if $\cF^\dagger$ is a Kraus so that $\cF(I)=I$ and $\cF^\dagger(I)=I$, then for any unitary matrices $U_0,U_T$,
\begin{equation}\label{eq:nonuniqueness}
U_T\cF^\dagger(U_0(\cdot)U_0^\dagger)U_T^\dagger
\end{equation}
is also a doubly stochastic Kraus map. {It is easy to see that the totality of doubly stochastic Kraus maps that relate to $\cE^\dagger$ via a multiplicative transformation are of this form.}

\subsection{The quantum bridge for general marginals and a conjecture}\label{conjecture}

Consider now the situation of the previous section with general initial and final density matrices $\rho_0$ and $\rho_T$. We are namely seeking a quantum bridge, as defined in Section \ref{sec:quantumbridge},  for the triple $(\cE_{0:T}^\dagger,\rho_0,\rho_T)$ . Theorem \ref{thm:doublystochastic0} admits the following generalization:
\begin{thm}\label{quantumbridge}
Given a  Kraus map $\cE_{0:T}^\dagger$ and two density matrices $\rho_0$ and $\rho_T$, suppose there exist observables
$\phi_0,\,\phi_T,\,\hat\phi_0,\,\hat\phi_T\in\fH_{++}$ solving the Schr\"{o}dinger system
\begin{eqnarray}\label{SchrSyst1}
\cE_{0:T}(\phi_T) &=& \phi_0,\\
\cE_{0:T}^\dagger(\hat{\phi}_0) &=& \hat{\phi}_T,\label{SchrSyst2}\\\rho_0&=&\x_0 \hat\phi_0\x_0^\dagger,\label{SchrSyst3a}\\
\rho_T&=&\x_T \hat\phi_T \x_T^\dagger.\label{SchrSyst3b}
\end{eqnarray}
Then, for any factorization
\begin{eqnarray*}
\phi_0 &=& \x_0^\dagger \x_0,\mbox{ and}\\ 
\phi_T &=& \x_T^\dagger \x_T,
\end{eqnarray*}
the map $\cF_{0:T}^\dagger:\fD\to\fD$ defined by
\begin{equation}\label{eq:doublystochastic2a}
\cF^\dagger(\cdot):= \x_T\left(\cE_{0:T}^\dagger(\x_0^{-1}(\cdot)\x_0^{-\dagger})\right)\x_T^\dagger
\end{equation}
is a quantum bridge for $(\cE_{0:T}^\dagger,\rho_0,\rho_T)$, namely $\cF(I)=I$ and $\cF^\dagger(\rho_0)=\rho_T$. 
\end{thm}

\noindent{\bf Proof:} 

$$
\cF_{0:T}(I)= \x_0^{-\dagger}\left[\cE_{0:T}\left(\x_T^{\dagger}(I)\x_T\right)\right]\x_0^{-1}=\x_0^{-\dagger}\left(\cE_{0:T}(\phi_T)\right)\x_0^{-1}=\x_0^{-\dagger}\left(\phi_0\right)\x_0^{-1}=I.
$$
Moreover,
$$
\cF_{0:T}^\dagger(\rho_0)= \x_T\left[\cE_{0:T}^\dagger\left(\x_0^{-1}\rho_0\x_0^{-\dagger}\right)\right]\x_T^\dagger=\x_T\left[\cE_{0:T}^\dagger(\hat\phi_0)\right]\x_T^\dagger=\x_T\left(\hat\phi_T\right)\x_T^\dagger=\rho_T.
$$
\hfill $\Box$

\noindent
The case of  initial and final pure states is worthwhile writing out.
\begin{cor}\label{purestates}
Given a positivity improving Kraus map $\cE_{0:T}^\dagger$ and two pure states
\[
\rho_0=v_0v_0^\dagger \mbox{ and } \rho_T=v_Tv_T^\dagger
\]
(i.e., $v_0,v_T$ are unit norm vectors),
define
\begin{align*}
\phi_0&:=\cE(v_Tv_T^\dagger)\\
\phi_T&:=v_Tv_T^\dagger,
\end{align*}
and
\[
\cF^\dagger(\cdot):= \phi_T^{1/2}\cE^\dagger(\phi_0^{-1/2}(\cdot)\phi_0^{-1/2})\phi_T^{1/2}
\]
(where, clearly, $\phi_T^{1/2}=\phi_T=v_Tv_T^\dagger$). Then, $\cF^\dagger$ is TPTP and satisfies the marginal conditions
\begin{align*}
\rho_T=\cF^\dagger(\rho_0).
\end{align*}
\end{cor}

\noindent{\bf Proof:} We readily verify that
\begin{align*}
\cF(I)&=\phi_0^{-1/2}\cE(\phi_T)\phi_0^{-1/2}\\
&=\phi_0^{-1/2}\cE(v_Tv_T^\dagger)\phi_0^{-1/2}\\
&=I.
\end{align*}
Next, we observe that
\begin{align*}
\cF^\dagger(\rho_0)&=\phi_T^{1/2}\cE^\dagger(\phi_0^{-1/2}(v_0v_0^\dagger)\phi_0^{-1/2})\phi_T^{1/2}\\
&=v_Tv_T^\dagger\cE^\dagger(\phi_0^{-1/2}(v_0v_0^\dagger)\phi_0^{-1/2})v_Tv_T^\dagger\\
&=v_Tv_T^\dagger\\
&=\rho_T.
\end{align*}
To see this, consider a representation $\cE^\dagger(\cdot)=\sum_i E_i(\cdot)E_i^\dagger$ and note that
\begin{align*}
v_T^\dagger\cE^\dagger(\phi_0^{-1/2}(v_0v_0^\dagger)\phi_0^{-1/2})v_T&=\sum_i
\left(v_T^\dagger E_i \phi_0^{-1/2}v_0\right)^2\\
&=v_0^\dagger\phi_0^{-1/2}\cE(v_Tv_T^\dagger)\phi_0^{-1/2}v_0\\
&=v_0^\dagger I v_0\\
&=1.
\end{align*}
\hfill $\Box$

Thus, precisely as in the classical case, the key challenge is to establish existence and uniqueness for the Schr\"{o}dinger system (\ref{SchrSyst1}-\ref{SchrSyst3b}). At present, proving the natural generalization of Theorem \ref{thm:doublystochastic} appears nontrivial.
Thus, below, we give the relevant statement as a conjecture since its proof remains elusive.

\begin{conj}\label{conj:doublystochastic}
Given a positivity-improving Kraus map $\cE_{0:T}^\dagger$, i.e., a Kraus map satisfying \eqref{eq:strictpositivity}, and given two density matrices $\rho_0$ and $\rho_T$, there exist observables $\phi_0$, $\phi_T$, $\hat{\phi}_0$, $\hat{\phi}_T$  in $\fH_{++}$ such that
\begin{align*}
\cE_{0:T}(\phi_T) &= \phi_0\\
\cE_{0:T}^\dagger(\hat{\phi}_0) &= \hat{\phi}_T
\end{align*}
with
\begin{subequations}
\begin{align}\label{eq:conj:a}
\rho_0&=\x_0 \hat\phi_0\x_0^\dagger\\
\rho_T&=\x_T \hat\phi_T \x_T^\dagger\label{eq:conj:b}
\end{align}
and
\begin{align}\label{eq:conj:c}
\phi_0 &= \x_0^\dagger \x_0\\
\phi_T &= \x_T^\dagger \x_T.\label{eq:conj:d}
\end{align}
\end{subequations}
In particular, $\x_0,\x_T$ can be taken to be 
Hermitian, i.e., for $i\in\{0,T\}$, $\x_i=(\phi_i)^{1/2}$ is the Hermitian square root of $\phi_i$.
\end{conj}

Equations \eqref{eq:conj:a} and \eqref{eq:conj:c} together, represent a non-commutative analogue of the relationship between $\hat\varphi(0,x_0),{\mathbf p}_0(x_0),\varphi(0,x_0)$ in
\[
\hat\varphi(0,x_0)= \frac{{\mathbf p}_0(x_0)}{\varphi(0,x_0)},
\]
while equations \eqref{eq:conj:b} and \eqref{eq:conj:d} represent the analogue of
\[
\hat\varphi(T,x_T)= \frac{{\mathbf p}_T(x_T)}{\varphi(T,x_T)}.\]
By taking $\x_0=(\phi_0)^{1/2}$, i.e., the Hermitian square root, clearly
\[
\hat\phi_0= (\phi_0)^{1/2}\rho_0(\phi_0)^{1/2}.
\]
On the other, taking $\x_T$ to be the Hermitian square root of $\phi_T$ and solving for $\phi_T$ in terms of $\hat\phi_T$ and $\rho_T$ using (\ref{eq:conj:b}-\ref{eq:conj:d}) gives
\[
\hat\phi_T\mapsto \phi_T=\left(
\rho_T^{1/2}\left(\rho_T^{-1/2}\hat\phi^{-1}\rho_T^{-1/2}
\right)^{1/2}\rho_T^{1/2}
\right)^2
\]

Thus, the conjecture claims the validity of the following correspondence,
\[
\begin{array}{ccccccc}
\rho_0&
\overset{\x_0^{-1}(\cdot)\x_0^{-\dagger}
}{\longrightarrow}&\rhoh & \overset{\cE_{0:T}^\dagger}{\longrightarrow} & \hat\phi_T & \overset{\x_T(\cdot)\x_T^{\dagger}
}{\longrightarrow} & \rho_T \\\\
\\
I & \overset{\x_0^{-\dagger}(\cdot)\x_0^{-1}}{\longleftarrow} &\phi_0=\x_0^\dagger \x_0 &\overset{\cE_{0:T}}{\longleftarrow} & \phi_T=\x_T^\dagger \x_T &\overset{\x_T^{\dagger}(\cdot)\x_T
}{\longleftarrow} & I
\end{array}
\]
and therefore, that the Kraus map
\[
\cF_{0:T}^\dagger(\cdot):=\x_T\left(\cE_{0:T}^\dagger(\x_0^{-1}(\cdot)\x_0^{-\dagger})\right)\x_T^{\dagger}
\]
solves the one-step quantum Sch\"odinger bridge problem for general marginal density matrices.
Extensive simulations have convinced the authors that the composition of maps
\begin{equation}\label{eq:composition:quantum}
\hat\phi_0  \overset{\cE_{0:T}^\dagger}{\longrightarrow} \hat\phi_T\overset{D_{T}}{\longrightarrow}\phi_T
\overset{\cE_{0:T}}{\longrightarrow} \phi_0
\overset{{\hat D}_0}{\longrightarrow}\left(\hat\phi_0\right)_{\rm next}
\end{equation}
where
\begin{eqnarray*}
D_T&:& \hat\phi_T\mapsto \phi_T=\left(
\rho_T^{1/2}\left(\rho_T^{-1/2}\hat\phi^{-1}\rho_T^{-1/2}
\right)^{1/2}\rho_T^{1/2}
\right)^2\\
\hat{D}_0&:& \phi_0\mapsto \hat\phi_0=(\phi_0)^{1/2}\rho (\phi_0)^{1/2}
\end{eqnarray*}
has an attractive fixed point \footnote{In this, $\x_i$ for $i\in\{0,T\}$ are taken Hermitian for specificity. Simulation shows that the iteration converges to a fixed point for a variety of other normalizations for the factors $\x_i$ of $\phi_i$, as well as when the boundary conditions are replaced by $\rho_0=(\hat\phi_0)^{1/2}\phi_0(\hat\phi_0)^{1/2}$ and $\rho_T=(\hat\phi_T)^{1/2}\phi_T(\hat\phi_T)^{1/2}$ and the iteration is modified accordingly.}. Unfortunately, $\hat D_0$ and $D_T$ are not isometries in the Hilbert metric as their commutative analogues.

{For a bridge with many intermediary time steps, very much as in the commutative classical case, the solution of the one-step bridge via multiplicative functional transformation permits solving the general bridge in a similar manner.} More specifically, starting from $\hat\phi_0,\phi_T$ that correspond to a fixed point of \eqref{eq:composition:quantum}, define for $i\in\{1,\ldots,T\}$
\begin{align*}
\hat\phi_i &:= \cE_{i-1}^\dagger\hat\phi_{i-1}\\
\phi_{i-1} &:= \cE_{i-1}\phi_i\\
\rho_i &:= (\phi_i)^{1/2}\hat\phi_i(\phi_i)^{1/2}.
\end{align*}
{The sought sequence of Kraus maps is
\[
\cF_{i+1}^\dagger(\cdot)=(\phi_{i+1})^{1/2}\left(\cE_i^\dagger(  (\phi_i)^{-1/2}(\cdot)(\phi_i)^{-1/2}  )\right)(\phi_{i+1})^{1/2},
\]
since, clearly,
\[
\cF_{0:T}^\dagger(\cdot)=(\phi_{T})^{1/2}\left(\cE_{0:T}^\dagger(  (\phi_0)^{-1/2}(\cdot)(\phi_0)^{-1/2}  )\right)(\phi_{T})^{1/2}
\]
and, assuming the validity of the conjecture, satisfies
\begin{eqnarray*}
\cF_{0:T}^\dagger(\rho_0)&=& (\phi_{T})^{1/2}\left(\cE_{0:T}^\dagger(  \hat\phi_0 )\right)(\phi_{T})^{1/2}\\
&=&(\phi_{T})^{1/2} \hat\phi_T (\phi_{T})^{1/2}\\
&=&\rho_T.
\end{eqnarray*}}

\section{Examples of doubly stochastic Kraus maps}
Perhaps the simplest nontrivial example of a (self-adjoint) positivity improving doubly stochastic Kraus map is
\begin{equation}\label{ex:kraus}
\cE^\dagger(\cdot)=E_1(\cdot)E_1^\dagger+E_2(\cdot)E_2^\dagger+E_3(\cdot)E_3^\dagger
\end{equation}
with
\[
E_1=\left[\begin{matrix}\sqrt{\frac{1}{2}} &0\\0&0\end{matrix}\right],\;
E_2=\left[\begin{matrix}0 &0\\0&\sqrt{\frac{1}{2}}\end{matrix}\right],\;
E_3=\left[\begin{matrix}0 &\sqrt{\frac{1}{2}}\\\sqrt{\frac{1}{2}}&0\end{matrix}\right].
\]
A second example is
\[
E_1=\left[\begin{matrix}\phantom{-}\sqrt{\frac{2}{3}} & \phantom{xxxx}0\\ \phantom{xxx}0 & \phantom{xxxx}0\end{matrix}\right],\;
E_2=\left[\begin{matrix}\phantom{-}\sqrt{\frac{1}{24}} & -\sqrt{\frac{1}{8}}\\ -\sqrt{\frac{1}{8}} & \phantom{-}\sqrt{\frac{3}{8}}\end{matrix}\right],\;
E_3=\left[\begin{matrix}\phantom{-}\sqrt{\frac{1}{24}} & \phantom{-}\sqrt{\frac{1}{8}}\\ \phantom{-}\sqrt{\frac{1}{8}} & \phantom{-}\sqrt{\frac{3}{8}}\end{matrix}\right]
\]
where all coefficient matrices are again symmetric but they all now of rank one.

In general, neither the symmetry (Hermitian-ness) of the coefficients nor any constraint on the rank is essential. The following example is constructed numerically. For this, we
start with
\begin{eqnarray*}
E_1&=&\left[\begin{matrix}1 & 1\\ 0 & 0\end{matrix}\right] M^{-1/2}\\
E_2&=&\left[\begin{matrix}0 & 1 \\ 0 & 1\end{matrix}\right] M^{-1/2}\\
E_2&=&\left[\begin{matrix}0 & 1 \\ 1 & 0\end{matrix}\right]M^{-1/2}
\end{eqnarray*}
where
\[M=\left[\begin{matrix}2&1\\1&4\end{matrix}\right]
\]
so that $E_1'E_1+E_2'E_2+E_3'E_3=I$.
It can be shown that the corresponding Kraus map $\cE^\dagger$ is positivity improving, i.e., it satisfies \eqref{eq:strictpositivity}.
We then compute the fixed point of \eqref{eq:contractivecomposition}, and this is
\begin{eqnarray*}
\phi_0 &=& 
\left[\begin{matrix}\phantom{-}1.1448 & -0.1350\\ -0.1350 & \phantom{-}0.8749\end{matrix}\right]\\
\phi_1 &=& 
\left[\begin{matrix}\phantom{-}0.8411 & -0.2362\\ -0.2362 & \phantom{-}1.3134\end{matrix}\right]\\
\hat\phi_0 &=& 
\left[\begin{matrix}\phantom{-}0.8897 & \phantom{-}0.1372\\ \phantom{-}0.1372 & \phantom{-}1.1642\end{matrix}\right]\\
\hat\phi_1 &=& 
\left[\begin{matrix}\phantom{-}1.2521 & \phantom{-}0.2251\\ \phantom{-}0.2251 & \phantom{-}0.8018\end{matrix}\right].
\end{eqnarray*}
These space-time harmonics give rise to coefficients
\begin{eqnarray*}
F_1&=&\left[\begin{matrix}\phantom{-}0.5690 & \phantom{-}0.4411\\ -0.0720 & -0.0558\end{matrix}\right]\\
F_2&=&\left[\begin{matrix}-0.0558 & \phantom{-}0.4411 \\ -0.0720 & \phantom{-}0.5690\end{matrix}\right]\\
F_3&=&\left[\begin{matrix}-0.1441 & \phantom{-}0.5131 \\ \phantom{-}0.8013 & -0.1441\end{matrix}\right]
\end{eqnarray*}
for a corresponding doubly stochastic Kraus map.

When $\rho_0$ and/or $\rho_T$ are in general different from the identity, extensive numerical experimentation suggests the validity of  Conjecture \ref{conj:doublystochastic} which, in conjuction with Theorem \ref{quantumbridge}, provides
solutions of the quantum Schr\"odinger bridge problem. In particular, for our first example \eqref{ex:kraus} and nonuniform marginals
\[\rho_0=\left[\begin{matrix}1/4 & 0\\ 0& 3/4\end{matrix}\right] \mbox{ and }
\rho_1=\left[\begin{matrix}2/3 & 0\\ 0& 1/3\end{matrix}\right]
\]
by iterating \eqref{eq:composition:quantum} we obtain 
\begin{eqnarray*}
\phi_0 &=& 
\left[\begin{matrix}1/2 & 0\\ 0 & 1/2\end{matrix}\right]\\
\phi_1 &=& 
\left[\begin{matrix}2/3 & 0\\ 0 & 1/3\end{matrix}\right]\\
\hat\phi_0 &=& 
\left[\begin{matrix}1/2 & 0\\ 0 & 3/2\end{matrix}\right]\\
\hat\phi_1 &=& 
\left[\begin{matrix}1 & 0\\ 0 & 1\end{matrix}\right]
\end{eqnarray*}
and
the Kraus map with coefficients
\begin{eqnarray*}
F_1&=&\left[\begin{matrix}\sqrt{2/3} & \phantom{xx}0\phantom{xx}\\\phantom{xx}0\phantom{xx}  & \phantom{xx}0\phantom{xx}\end{matrix}\right]\\
F_2&=&\left[\begin{matrix}\phantom{xx}0\phantom{xx} &\phantom{xx}0\phantom{xx}  \\ \phantom{xx}0\phantom{xx}& \sqrt{1/3}\end{matrix}\right]\\
F_3&=&\left[\begin{matrix}  \phantom{xx}0\phantom{xx}& \sqrt{2/3} \\ \sqrt{1/3} & \phantom{xx}0\phantom{xx} \end{matrix}\right].
\end{eqnarray*}
This Kraus map is no longer Hermitian, it is of the form \eqref{eq:doublystochastic2} and, as can be readily verified, it satisfies the required condition $\cF^\dagger(\rho_0)=\rho_1$.
Software for numerical experimentation can be found at \url{http://www.ece.umn.edu/~georgiou/papers/schrodinger_bridge/}.

\section{Concluding remarks}

{In this paper, we introduced a new approach to studying Schr\"{o}dinger's systems.
In particular, we establish new proofs for existence and uniqueness of solutions.
In contrast to earlier treatments, our approach provides a direct computational procedure for obtaining the space-time harmonic function and the corresponding solution of the Schr\"odinger bridge in finite-dimensions. Space-time harmonics are obtained as fixed points of a certain map. Convergence is established in a suitable {\em projective geometry} as convergence of {\em rays} using the Hilbert metric. That is, in the classical case of discrete random vectors that is treated herein, our approach provides a direct new proof of existence and uniqueness by a contraction mapping principle. Since the approach also provides a computational scheme, it appears to have considerable potential for applications. Indeed, models for stochastic evolution, which include Markov chains, are ubiquitous. The bridge evolution may be viewed as a controlled  {\em steering} problem --a facet that will be thoroughly explored elsewhere. In the case of quantum channels, the solution of an analogous Schr\"{o}dinger system corresponds to a steering between two given density matrices. We prove convergence of an iterative algorithm in a corresponding projective metric in the case of uniform marginals, i.e., identity matrices, thereby establishing existence of doubly stochastic Kraus maps that can be derived from a given reference map via multiplicative functional transformations. Extensive simulations have convinced the authors of a more general result regarding the general quantum bridge problem which, however, is stated as a conjecture since at present a proof is not available.}

\section*{Acknowledgment}
The authors are very grateful to Yongxin Chen for a correction in the proof of Theorems 3 and 6.
\bibliographystyle{siam}
\bibliography{bibliography}
\end{document}